\documentstyle[12pt]{article}
\vspace{1 cm}
\def\lapproxeq{\lower .7ex\hbox{$\;\stackrel{\textstyle <}{\sim}\;$}}
\def\gapproxeq{\lower .7ex\hbox{$\;\stackrel{\textstyle >}{\sim}\;$}}

\begin{document}

\titlepage


\begin{center}
\vspace*{2cm}
{\large{\bf Analytic Approach to Small $x$ Structure Functions}}
\end{center}

\vspace*{.75cm}
\begin{center}
J.R.\ Forshaw, R.G.\ Roberts and R.S.Thorne\\
Rutherford Appleton Laboratory, \\ Chilton, Didcot OX11 0QX, England. \\
\end{center}

\vspace*{1.5cm}

\begin{abstract}
We present a method for the analytic solution of small $x$ structure functions.
The essential small $x$ logarithms are summed to all orders in the anomalous
dimensions and coefficient functions. Although we work at leading logarithmic
accuracy, the method is general enough to allow the systematic inclusion of
sub-leading logarithms. Results and predictions are presented for the gluon
density, and the structure functions $F_2(x,Q^2)$ and $F_L(x,Q^2)$.
We find that corrections to the simple
double logarithmic calculation are important in the HERA range and obtain good
fits to all available data.
\end{abstract}

\newpage
As a result of the recent work of Catani and Hautmann \cite{ch},
it is now possible to
include the dominant small $x$ dynamics encompassed by the formalism of
Balitsky, Fadin, Kuraev and Lipatov (BFKL) \cite{bfkl}
within the framework of the renormalisation group and collinear factorisation,
and some (mostly numerical) studies have already been performed \cite{bf1,ehw}.
In this paper, we wish to
present an analytic solution to the relevant evolution equations and their
convolution with the appropriate coefficient functions. Throughout we work in
the high energy limit, i.e. we sum all terms in the perturbative expansion of
the cross section which are
$$ \sim \left( \alpha_s \ln \frac{s}{Q^2}\right)^n, $$
where $s$ is the relevant centre-of-mass energy and $Q^2$ characterises the
typical short distances involved. We shall focus on deep inelastic scattering
at the DESY $ep$ collider, HERA. In which case, $\sqrt{s}$ is the $\gamma p$
centre-of-mass energy and $-Q^2$ is the photon virtuality, i.e. the Bjorken-$x$
variable, $x \approx Q^2/s$. Our approach is quite general and it will be
clear how to extend it beyond the leading logarithmic accuracy.

\section*{Altarelli-Parisi Evolution at small $x$ and the gluon density}
Recall the Dokshitzer, Gribov, Lipatov, Altarelli, Parisi (DGLAP) equations for
the parton distribution functions \cite{dglap}:
\begin{equation}
\frac{\partial f_N^i(Q^2)}{\partial \ln Q^2} = \sum_j \gamma_N^{ij}
f_N^j(Q^2).
\end{equation}
$f_N^i(Q^2)$ is the Nth moment of the momentum distribution for partons
of type $i$ and $\gamma_N^{ij}$ is the anomalous dimension matrix, i.e.
\begin{eqnarray}
f_N^i(Q^2) &=& \int_0^1 dx \, x^{N-1} f^i(x,Q^2), \nonumber \\
\gamma_N^{ij} &=& \int_0^1 dx \, x^N P_{ij}(x).
\end{eqnarray}
Our notation is such that the important gluon anomalous dimension,
\begin{equation}
 \gamma_N^{gg} = \frac{\bar{\alpha}_s}{N} + 2 \zeta(3)\left(
\frac{\bar{\alpha}_s}{N}\right)^4 + ....
\end{equation}
and $\bar{\alpha}_s = 3\alpha_s/\pi$.

These equations are solved given the boundary conditions, $f_N^i(Q_0^2)$, i.e.
they allow the $Q^2$-dependence of the parton distribution functions to be
determined but not their absolute normalisation.

In the high energy (i.e. small $x$) limit, we keep only those terms in the
anomalous dimension matrix which are $\sim (\alpha_s/N)^n$, i.e. the leading
logarithmic terms in the splitting functions. In this case, evolution is driven
by $\gamma_N^{gg}$, which satisfies \cite{bfkl,j}
\begin{equation}
1 = \frac{\bar{\alpha}_s}{N} \chi(\gamma_N^{gg}),
\end{equation}
where
\begin{equation}
\chi(\gamma) = 2 \psi(1) - \psi(\gamma) - \psi(1-\gamma)
\end{equation}
and $\psi(\gamma)$ is the Euler-gamma function. The first two non-zero
terms in the series expansion are written in eq.(3). The DGLAP equations then
have the simple solution:
\begin{eqnarray}
f_N^s(Q^2) &=& f_N^s(Q_0^2), \nonumber \\
f_N^g(Q^2) &=& \left[ f_N^g(Q_0^2) + \frac{4}{9} f_N^s(Q_0^2) \right]
\exp\left( \int_{Q_0^2}^{Q^2} \frac{dk^2}{k^2} \gamma_N^{gg} \right) -
\frac{4}{9} f_N^s(Q_0^2).
\end{eqnarray}
The singlet quark density is $f_N^s(Q^2) = \sum_i f_N^i(Q^2)$ where the sum
runs over all quarks and anti-quarks. Since we work in the small $x$ region, we
expect the gluon density to be dominant and subsequently drop all reference to
the singlet density (except implicitly in the input to  $F_{2}(x,Q^2)$).
We have explicitly checked that this makes very little
quantitative difference to our results.

In order to construct a sensible gluon structure function, we do not merely
invert the $N$-space solution above. It is more natural to define the gluon
structure function to be that object which would be observed if we had a
coloured current available as our probe. In which case there are important
contributions which arise, not only from the QCD evolution but also from the
coefficient function. One can think of such corrections as arising from graphs
which should not be exponentiated via the renormalisation group and so
contain no explicit strong ordering of the rung momenta. These graphs are
essential for a sensible definition of the gluon structure function (e.g. as
the object which is closely related to the structure function $F_L(x,Q^2)$)
and for consistency with the structure function which is constructed by
integrating the `unintegrated gluon density', ${\cal{F}}(x,k^2)$, obtained by
solving the BFKL equation \cite{cch}.
To see this, we start from the BFKL definition of the gluon structure function,
i.e.
\begin{equation}
G(x,Q^2) = \int_0^{\infty} \frac{dk^2}{k^2} \int_x^1 \frac{dz}{z} \delta(1-x/z)
\Theta(Q^2 - k^2) {\cal{F}}(z,k^2).
\end{equation}
The hard scatter cross section which is to be convoluted with
${\cal{F}}(x,k^2)$ is thus
\begin{equation}
\hat{\sigma}_N(k^2/Q^2) = \int_0^1 dx x^{N-1} \delta(1-x) \Theta(Q^2-k^2).
\end{equation}
Now, in the formalism of Catani and Hautmann \cite{ch},
\begin{equation}
G_N(Q^2) = \gamma_N^{gg} \int_0^{\infty} \frac{dk^2}{k^2} \left(
\frac{k^2}{Q^2}\right)^{\gamma_N^{gg}} \hat{\sigma}_N(k^2/Q^2) \; R_N \;
f_N^g(Q^2),
\end{equation}
i.e.
\begin{equation}
G_N(Q^2) = R_N \; f_N^g(Q^2).
\end{equation}
Here $R_N$ plays the role of a coefficient function for the `natural'
definition of the gluon structure function \cite{cch,hq}.
This important factor, which is a process independent but scheme
{\it dependent} quantity
is given, in the $\overline{ {\mathrm{MS}} }$ scheme (which we shall employ
throughout this paper) by
\begin{eqnarray}
R_N = & & \left\{ \frac{ \Gamma(1-\gamma_N^{gg})\; \chi(\gamma_N^{gg})}
{\Gamma(1+\gamma_N^{gg})\; [-\gamma_N^{gg}\;
\chi'(\gamma_N^{gg})]} \right\}^{1/2} \times \nonumber \\
& & \exp \left\{ \gamma_N^{gg} \psi(1) + \int_0^{\gamma_N^{gg}} d\gamma
\frac{\psi'(\gamma)-\psi'(1-\gamma)}{\chi(\gamma)} \right\}.
\end{eqnarray}

\section*{Solution in $x$-space}
Let us now show how to invert the solution for $G_N(Q^2)$ back into $x$-space.
We must perform the integral
\begin{equation}
G(x,Q^2) =\frac{1}{2\pi i} \int dN \, f^g_N(Q_0^2) \; R_N \; \exp\left( N \ln
1/x
 + Z_N(Q^2) \right)
\end{equation}
where the integral is over a contour to the right of all singularities, and
\begin{eqnarray}
Z_N(Q^2) &=& \int_{Q_0^2}^{Q^2} \frac{dk^2}{k^2} \gamma_N^{gg} \nonumber \\
         &=& \frac{\gamma^2 \zeta}{N} + \sum_{m=2}^{\infty} \frac{a_m}{m-1}
\frac{\gamma^2}{N} \left[ \left(\frac{\bar{\alpha}_s(Q_0^2)}{N}\right)^{m-1} -
\left(\frac{\bar{\alpha}_s(Q^2)}{N}\right)^{m-1}\right] \nonumber \\
         &=& \sum_{m=1}^{\infty} \frac{b_m(\zeta)}{N^m}.
\end{eqnarray}
$\zeta = \ln (\alpha_s(Q_0^2)/\alpha_s(Q^2))$, $\gamma^2 = 12/\beta_0$ and,
for consistency with the standard approach, we run the coupling at the scale
$k^2$ (in the anomalous dimension integral). However, we note that at the
leading logarithmic accuracy this is an essentially free choice.
The coefficients $a_n$ define the series expansion of the gluon anomalous
dimension:
\begin{equation}
\gamma_N^{gg} = \sum_{n=1}^{\infty} a_n \left(
\frac{\bar{\alpha}_s}{N}\right)^n.
\end{equation}
We write the factor $R_N$ also as a series expansion:
\begin{equation}
R_N = \sum_{n=0}^{\infty} c_n \left(
\frac{1}{N}\right)^n,
\end{equation}
and we note that $R_N = 1 + \frac{8}{3} \zeta(3) (\bar{\alpha}_s/N)^3 + ...$.

We choose the boundary condition,
$$ f^g_N(x, Q_0^2) = {\cal{N}} \Theta(x_0 - x),$$ which becomes $f^g_N(Q_0^2)
={\cal{N}}/N$ in moment space if we choose $x/x_0$ as the conjugate variable
to $N$. This starting distribution is a good approximation to the expected
$\sim (1-x)^5$ behaviour if $x_0 \sim 0.1$, and leads
to reliable results for $x \lapproxeq 0.01$. The choice of a flat
starting distribution (at small $x$) is motivated by the known behaviour
of total cross sections at high energies, i.e. the `soft' pomeron is known to
have intercept close to 1 \cite{pom}. It is the
small $x$ behaviour one would expect in
the absence of any perturbative QCD corrections.

We can now perform the $N$-plane integral by taking a particular choice of
contour to be the line from $r-i\infty$ to $r+i\infty$ plus a circle with
centre at the origin and radius $r>4 \bar \alpha_s (Q_0^2)\ln2$. This choice
of contour ensures the analyticity of the integrand along the contour, i.e.
that all cuts lie within the circle (e.g. for the precise location of
the cuts see ref.\cite{ehw}).
The value of the integral is now equal to that over the circle, and
putting $N = r e^{i \theta}$ we obtain in a straightforward manner:
\begin{eqnarray}
G(x,Q^2) &=& {\cal{N}}\sum_{i=0}^{\infty} c_i \left(
\frac{\xi}{\gamma^2 \zeta}
\right)^{i/2} I_i(2 \gamma \sqrt{\xi \zeta}) \nonumber \\
         &+& {\cal{N}}\sum_{i=0}^{\infty} \sum_{j=2}^{\infty} c_i b_j
\left( \frac{\xi}{\gamma^2 \zeta} \right)^{(i+j)/2}
I_{i+j}(2\gamma \sqrt{\xi \zeta}) \nonumber \\
         &+& {\cal{N}}\sum_{i=0}^{\infty} \sum_{j=2}^{\infty}
\sum_{k=2}^{\infty} \frac{c_i b_j b_k}{2} \left( \frac{\xi}{\gamma^2 \zeta}
\right)^{(i+j+k)/2} I_{i+j+k}(2\gamma\sqrt{\xi \zeta}) + ........
\end{eqnarray}
where $\xi = \ln (x_0/x)$. We note that exactly the same method could be used
if we were to choose a powerlike input, or even the $(1-x)^5$ behaviour. One
simply finds the moment space expression for the input and expands in powers
of $N$. We also note that for small $x$ the result obtained using the
saddle-point method to evaluate eq.(12) does not give a good approximation and
provides misleading results. This failure occurs essentially because, along
the contour of steepest decent, the integrand does
not fall quickly enough for values of $N$ far from the saddle-point.

Let us now discuss our solution. Firstly, we see explicitly the double log
result, $$ G(x,Q^2) \sim I_0(2 \gamma \sqrt{\xi \zeta}),$$
which arises when only the leading order (in $\alpha_s$) terms are kept.
Going beyond this first term,
the inverse factorials associated with the Bessel functions ensure that
the summations in our expression converge for all $x$,
despite the fact that the expansions of $\gamma^{gg}_N$ and $R_{N}$ diverge
for $N<4  \bar \alpha_s\ln 2$. This effect of convergence in $x$-space
was pointed out using a similar, but slightly less direct,
argument in \cite{bf1}. Using the first sum in eq.(16)
we can recover the power behaviour of the structure function at small
enough $x$. It arises after taking the small argument expansion of the Bessel
function (which is appropriate for large order Bessel functions) and using the
fact that $c_{n+1}/c_n = 4 \bar{\alpha}_s \ln 2$ for asymptotically large $n$,
i.e. the general term in the sum over $i$ is
$$ \sim  (4 \bar{\alpha}_s \ln 2)^i \left( \frac{\xi}{\gamma^2 \zeta}
\right)^{i/2} \frac{(\gamma^2 \zeta \xi)^{i/2}}{i\,{\rm !}} = \frac{
( 4 \xi\bar{\alpha}_s \ln 2)^i}{i\,{\rm !}}. $$
A similar power behaviour is generated by the other terms in eq.(16), i.e. due
to the $Q^2$-evolution, but they are not important for all practical values of
$x$. The same qualitative conclusion has been reached in ref.\cite{bf1}, where
a sum of Bessel functions was presented as an approximate solution for
$f^g(x,Q^2)$. In other words,
the dominant corrections to the double log result are due
to the presence of the $R_N$ factor, i.e. the
corrections to the evolution are small (due to the relatively small
size of the coefficients in the expansion of the gluon anomalous dimension),
only becoming dominant at very large $Q^2$ and/or very small $x$.

The fraction of $G(x,Q^2)$ which arises solely from the double log graphs
(i.e. the $I_0$ Bessel function) is presented in the contour plot shown in
fig.(1). It can be seen that the high energy (BFKL) corrections are significant
over the HERA range despite the fact that the coefficients $c_1$ and $c_2$
vanish. We note that the contribution from the BFKL corrections to the
evolution (i.e. those terms involving the $b_i$ coefficients) are almost
entirely negligible, in fact they contribute less than $4\%$ over the
$x$-$Q^2$ range probed at HERA. In fig.(2), we show the $x$ dependence of
$G(x,Q^2)$ at different $Q^2$ values and compare to the double log
contribution. In all of our plots, we choose $x_0=0.1$ and
take ${\cal N} = 1.1$ and $Q_0^2 =
2.0$. ${\cal N}$ and $Q_0^2$ are the only parameters for the gluon,
and are fixed by fitting $F_2(x,Q^2)$ to
the HERA data (see the following section for a discussion of this procedure).
Note that our approach does not permit a flat gluon structure function,
even though our input density was flat. This is in keeping with the standard
BFKL result developed by direct solution of the BFKL equation. The
scale  $Q_0$ is to be understood as the scale below which we cannot use the
perturbative approach. As such, we are unable to make any definite statements
regarding the eventual saturation and flattening off of the small $x$
structure functions (as $Q^2$ falls below $Q_0^2$) since this procedure is
governed by physics beyond that which is considered here. (However, we do see
a hint of the breakdown of our approach, as we will discuss in the next
section.) Indeed, we shall find that the quality of our fits to
$F_2(x,Q^2)$ is largely insensitive to the choice of $Q_0^2$ once it is above
$\approx 1$ GeV$^2$. Our approach should be contrasted to that which attempts
to evolve from some, typically quite low, value of $Q_0^2$ with a flat (or
valencelike) starting distribution to higher $Q^2$ \cite{grv,bf}.

\section*{Deep inelastic structure functions}
In the previous section we concentrated on the gluon structure function,
$G(x,Q^2)$. It involves no new techniques to extend the formalism to the
case of the deep
inelastic structure functions, $F_2(x,Q^2)$ and $F_L(x,Q^2)$. Eq.(10), which
defines the gluon structure function, is merely a specific form of the more
general expression for the dimensionless cross section, $F(x,Q^2)$:
\begin{equation}
F_N(Q^2) = C_N(Q^2/\mu_F^2) f^g_N(\mu_F^2)
\end{equation}
where $\mu_F^2$ is the factorisation scale (chosen to equal $Q^2$)
and $C_N(Q^2/\mu_F^2)$ is the coefficient function (equal to $R_N$ in the case
of the gluon structure function).

Catani and Hautmann have shown that the coefficient function can be factorised
into a product of the process independent (but factorisation scheme dependent)
factor, $R_N$, and a process dependent factor, $h_N(\gamma_N^{gg})$, where
\cite{ch}
\begin{equation}
h_N(\gamma) \equiv \gamma \int_0^{\infty} \frac{dk^2}{k^2} \left(
\frac{k^2}{\mu_F^2} \right)^{\gamma} \hat{\sigma}_N(k^2/Q^2).
\end{equation}
The hard subprocess cross section, $\hat{\sigma}_N(k^2/Q^2)$, is the lowest
order (in $\alpha_s$) cross section for scattering off-shell gluons (off the
virtual photon in the case of deep inelastic scattering).

Thus, for the structure function, $F_L(x,Q^2)$
\begin{equation}
F_{L,N}(Q^2) = \langle e_q^2 \, \rangle h_{L,N}(\gamma^{gg}_N)\, R_N \,
f_N^g(Q^2)
\end{equation}
with
\begin{equation}
h_{L,N}(\gamma) = \frac{\alpha_s}{2 \pi} N_f T_R \frac{4(1-\gamma)}{3 -
2\gamma} \frac{\Gamma^3(1-\gamma)
\Gamma^3(1+\gamma)}{\Gamma(2-2\gamma)\Gamma(2+2\gamma)}
\end{equation}
and where $\langle e_q^2 \rangle$ is the mean quark charge squared.
So, in order to evaluate $F_L(x,Q^2)$, we merely replace the $c_n$ coefficients
in eq.(16) by the corresponding coefficients in the expansion of
$h_{L,N}(\gamma_N^{gg}) \; R_N$.

Similarly,
\begin{equation}
h_{2,N}(\gamma) = \frac{2+3\gamma-3\gamma^2}{2(1-\gamma)} h_{L,N}
\end{equation}
determines the $Q^2$-dependence of $F_2(x,Q^2)$, i.e.
\begin{eqnarray}
\frac{\partial F_{2,N}(Q^2)}{\partial \ln Q^2} &=& \langle e_q^2\rangle
[ C^{(1)}_{g,N}(\alpha_s(Q^2)) \gamma_N^{gg} +2n_f\gamma_N^{qg} ]f^g_N(q^2)
\nonumber \\
         &=& \langle e_q^2 \rangle
\, h_{2,N}(\gamma_N^{gg})\, R_N\, f_N^g(Q^2),
\end{eqnarray}
to formally leading order.

In fig.(3), using the same choice of parameters (no more are needed) as in the
discussion of $G(x,Q^2)$ previously, we present our predictions for the
longitudinal structure function, $F_L(x,Q^2)$. As well as the full solution, we
show the double log contribution to it. The largeness of the corrections to the
double log calculation (in comparison to case of $G(x,Q^2)$), can be traced
back to the fact that the second and third coefficients in the expansion of
$h_{L,N} R_N$ are no longer zero. Also shown in fig.(3) is the result of
re-fitting the HERA data on $F_2(x,Q^2)$, while keeping only the double log
Bessel function. In order unambiguously to establish the existence of the
high energy corrections, it is ultimately necessary to expose deviations
from the double log approach (or more precisely approaches which do not
sum the infinity of high order corrections ${\cal{O}}(\alpha_s/N)$) and so
this is the reason for our comparison. As seen, the prediction from the
double log approach is mostly larger than that for the full expression, but
flatter with $x$. This largeness comes about
mainly because the starting scale is much lower, and hence
there has been more time for evolution to take place.

Let us now turn to the structure function, $F_2(x,Q^2)$ and its comparison with
the HERA data \cite{heradat}.
We start by considering the expression of eq.(22). In order
to construct $F_2(x,Q^2)$, we must integrate over $Q^2$ and invoke an input
distribution, $F_2(x,Q_0^2)$. We choose this to be of the form $A +
Bx^{-\lambda}$. We see no reason to believe that the input form of
$F_2(x,Q^2)$ should be purely flat since, as demonstrated, the gluon structure
function always has some powerlike behaviour due to the coefficient
function. Indeed, we are not able to obtain a very good fit with a
completely flat input. We could have chosen an input of $B x^{-\lambda}$, and
still obtain a comparable fit. However, our aim is not simply to obtain
the best fit with the least number of parameters, but to determine the
behaviour of the structure function as accurately as possible, and we believe
the chosen input is the best way to do this.
This introduces three extra parameters. The values of our 5 parameters
($Q_0$, the normalisation of the input gluon density and the
three parameters in $F_2(x,Q_0^2)$) are then obtained by fitting to the HERA
data. Throughout, we
work with 4 quark flavours and $\Lambda_{QCD} = 115$ MeV.  Agreement with
the data is
very good, i.e. $\chi^2 = 48$ for the 92 data points which have $x < 0.01$.

Although we obtain very good agreement with the available HERA data, we expect
our results to be subject to important corrections. Let us now explain why. In
the leading log approximation, the structure function
$F_2(x,Q^2)$ can be written in the form
\begin{eqnarray}
F_{2,N}(Q^2) &=& \langle e_q^2\rangle C^{(1)}_{g,N}(\alpha_s(Q^2))f^g_N(Q^2)
+ [C^{(0)}_{q,N} + C^{(1)}_{q,N}(\alpha_s(Q^2))] 2n_f \langle e_q^2 \rangle
f^s_N(Q_0^2) \nonumber\\
&+& 2n_f\langle e^2_q\rangle C^{(0)}_{q,N} \int_{Q_0^2}^{Q^2}
\frac{dq^2}{q^2} (\gamma_N^{qg}f^g_N(q^2) + \gamma_N^{qq}f^s_N(Q_0^2)).
\end{eqnarray}
The superscript on the coefficient functions specifies the order (in
$\alpha_s$) of the contribution, i.e. $C_{i,N} = \sum_n C^{(n)}_{i,N}$. The
parton model coefficient functions are $C^{(0)}_{g,N} = 0$ and
$C^{(0)}_{q,N} = 1$.

Taking the derivative of this expression leads to eq.(22), but only after
neglecting the higher order terms which are induced by differentiating the
coefficient functions. Such terms are formally sub-leading since
$$ \frac{\partial}{\partial \ln Q^2} = -\alpha_s^2 \frac{\beta_0}{4 \pi}
\frac{\partial}{\partial \alpha_s}.$$ However, they are not sub-leading once
eq.(22) is integrated to form the structure function, $F_2(x,Q^2)$.

To see how important these corrections are expected to be, we expand the
coefficient function
\begin{equation}
C^{(1)}_{g,N} = \sum_{n=0}^{\infty} p_n\; \bar{\alpha}_s \left(
\frac{\bar{\alpha}_s}{N} \right)^n.
\end{equation}
The ratio of the term $\sim \bar{\alpha}_s (\bar{\alpha}_s/N)^m$ in
$\partial C^{(1)}_{g,N}/\ln Q^2$ to the corresponding term in the series
expansion of $C^{(1)}_{g,N}\;\gamma_N^{gg}$ is thus
\begin{equation}
\left( - \frac{\beta_0 \alpha_s}{4 \pi} \right) \frac{(m+1)\;
b_m}{\sum_{i=0}^{m-1} b_{i+1} a_{m-i}}.
\end{equation}
Since $a_{n+1}/a_{n}= 4\ln 2$ for large $n$ and, assuming a similar
relation for the $p_n$ coefficients, it follows that this ratio becomes
$$ \sim -\frac{\beta_0 \alpha_s}{4\pi} \frac{1}{A} $$
where $a_{n} \approx A \;(4 \ln 2)^{n-1}$. Since $A \ll 1$ we cannot
ignore such
contributions. We should emphasise that $p_{n+1}/p_{n}$ cannot exceed $4 \ln 2$
(since we know the dominant singularity arises at $N = 4 \bar{\alpha}_s\ln 2 $)
and that assuming any $p_{n+1}/p_{n} < 4 \ln 2$ leads to an even stronger
enhancement of the derivative terms (e.g. by a factor of $m$ for $p_{n+1}/p_{n}
\ll 4 \ln 2$). All the evidence from the calculation of the series expansion
of the coefficient function is that $p_{n+1}/p_n$ is indeed $\sim 4\ln2 $ for
large $n$.

Although these corrections are formally sub-leading, we believe it is unlikely
that they will cancel with higher order graphs and as such it is safe (and
probably more appropriate) to include them in this order of the calculation.
Since they lead to a negative contribution to $\partial F_2(x,Q^2)/ \partial
\ln Q^2$, we therefore expect a reduction in $\partial
F_2(x,Q^2)/ \partial \ln Q^2$ as $x$ falls at fixed $Q^2$ (or $Q^2$ falls at
fixed and small $x$).

Using the expressions of ref.\cite{ch}
we have computed the series expansions of the gluon coefficient function,
$C_{g,N}^{(1)}$, and the anomalous dimension, $\gamma_N^{qg}$, in the
$\overline{ {\mathrm{MS}} }$ scheme to 18th order (we will shortly have more
to say on the choice of scheme), the first six terms of course agreeing
with the numerical values explicitly given in ref.\cite{ch}.
It is then a simple matter to calculate $F_2$ by transforming eq.(23)
(using eq.(6) to give $f_N^g(Q^2)$)
to $x$-space, in the same way that we obtained eq.(16) from eq.(12). We
make the simplification that (as in eq.(16), and as when
obtaining $F_L(x, Q^2)$ by transforming eq.(19) to $x$-space) we drop the
quark singlet density in the definition of the gluon distribution
(as explained in the remarks following eq.(6)).
We also neglect $\gamma^{qq}_N$ and the ${\cal O}(\alpha_s)$ contributions to
the quark coefficient function when calculating
$F_2(x,Q^2)$ (since the input quark density is small in
comparison to the evolved gluon density, the quark coefficient
function is smaller than that for the gluon and $\gamma^{qq}_N$ is
smaller than $\gamma^{qg}_N$). Using this method, we compute $F_2(x,Q^2)$
including those corrections which were neglected when eq.(22) was integrated
over $Q^2$. The solid line in fig.(4) shows the result of a new fit to the HERA
data and marginal improvement in the $\chi^2$ of the fit is found, i.e.
$\chi^2 = 45$ for the 92 data points. A considerable improvement in the
insensitivity to the  value of $Q_0^2$ is also found.
For our best fit $Q_0^2 = 2.0$ and ${\cal N}=1.1$,
and these are the parameters used to determine our predictions for $G(x,Q^2)$
and $F_L(x,Q^2)$. We also find the input to $F_2(x, Q^2)$ to be $0.15
+0.035 x^{-0.4}$.

{}From our results we conclude that our choice of a
$\Theta$-function form of the gluon input is appropriate.
Also, we note that the $\chi^2$ only starts to worsen significantly once
$Q_0^2 \lapproxeq 1$ GeV$^2$.
This is consistent with idea that the scale at which we define our input
should be essentially arbitrary, providing it is large enough for the
perturbative expansion to apply, and not too large to fill the available
phase space. The dotted line in the figure shows the previously discussed best
fit for $F_2(x,Q^2)$, i.e. ignoring the derivatives of the
coefficient function. The dashed line shows the result of fitting the data
taking only the leading term in the Bessel function expansion, i.e. the double
log result, and flat inputs for the gluon and for $F_2(x, Q^2)$. This also has
a very good $\chi^2$ of 44 for the 92 points, but high sensitivity to $Q_0^2$.
It is clear from the plots that the differences between the full leading
log calculation of $F_2(x,Q^2)$ and the dotted (and dashed) line
are consistent with our expectations.

As a slight word of caution we must note that for low enough $x$,
$\partial F_2(x,Q^2)/
\partial \ln Q^2$ eventually becomes negative, and $F_2(x,Q^2)$ rises
for falling $Q^2$. It is this region of negative $\ln Q^2$ derivative
where our calculation starts to become untrustworthy. However,
the effect only sets in for $x\sim 10^{-4}$ at $Q^2=Q^2_0$,
and the value of $x$ at which $\partial F_2(x, Q^2)/ \partial \ln Q^2 =0$
falls very rapidly as $Q^2$ increases.
Similarly, we feel that the predictions for the gluon
and $F_{L}(x,Q^2)$ should be viewed with a little caution at extremely low
$x$ and small $Q^2$ . Nevertheless, it is
reassuring that the region of breakdown is where we might
expect physics beyond that considered in our approach to become important.

It is also important to comment on our choice of scheme.
We could just as well have computed $F_2(x,Q^2)$ in the DIS scheme, and
obtained precisely the same results as in $\overline{ {\mathrm{MS}} }$
(see below). This is true providing we take care to include the sub-leading
corrections to $\gamma^{qg}_N$ which contribute in the leading order to
$F_2(x,Q^2)$. These terms are those neglected in eq.(5.27) of Catani and
Hautmann when transforming to DIS scheme.

Before concluding, we wish to make a few remarks regarding the recent
conclusions of Ellis {\it et al} \cite{ehw}. Recall that
$F_{2,N}(Q^2) - F_{2,N}(Q_0^2) \sim \alpha_s {\cal{O}}(\alpha_s/N)$,
and as such any calculation which sums
the leading logarithms must necessarily include the quark anomalous dimensions,
$\gamma_N^{qg}$ (and $\gamma_N^{qq})$, computed to the same accuracy (i.e.
$\sim \alpha_s {\cal{O}}(\alpha_s/N)$). Thus, it
is not sensible to compute $F_{2,N}(Q^2)$ by including only the BFKL anomalous
dimension, $\gamma_N^{gg} \sim {\cal{O}}(\alpha_s/N)$, and (for example) the
two-loop form for the rest of the anomalous dimension matrix. It is
also important to appreciate that, in the DIS scheme (where all the
gluon coefficient functions are zero) the quark anomalous dimension
$\gamma_N^{qg}$ is where the
physics which gave the dominant (leading log) contribution to the gluon
structure function $G(x,Q^2)$ resides, i.e.
\begin{equation}
2 n_f \; \gamma_N^{qg} = R_N \; h_{2,N}(\gamma_N^{gg})
\end{equation}
determines the DIS scheme anomalous dimension in terms of $R_N$ (calculated in
$\overline{{\mathrm{MS}}}$ scheme). This is in contrast to the
$\overline{{\mathrm{MS}}}$ scheme, where it is the coefficient function
$C^{(1)}_{g,N}$ which has the large coefficients in the series
expansion in $\bar\alpha_s/N$ (due to the corresponding
large coefficients in $R_N$) and, as a result, the coefficients of
$\gamma_N^{qg}$ are much smaller in the $\overline{{\mathrm{MS}}}$ scheme than
in the DIS scheme. Therefore, the relatively
small effect observed in ref.\cite{ehw} when only the BFKL anomalous
dimension is added to the full two-loop formalism is consistent with
our finding that the $b_i$ coefficients in eq.(16) have little
impact. However, on adding the all orders quark anomalous dimension,
$\gamma_N^{qg}$, Ellis {\it et al} found huge effects which are
very sensitive to how momentum conservation is applied. The largeness of their
effect is driven by the largeness of the coefficients in the expansion of
$\gamma_N^{qg}$ as calculated in the DIS scheme. We wish to urge caution in
interpreting their results as evidence of important higher order corrections to
the leading log BFKL framework. In particular, as we have just emphasised
the bulk of the BFKL physics resides in the leading log factor $R_N$ (i.e. in
the DIS scheme $\gamma_N^{qg}$), so part of the large effect which is
seen after turning on $\gamma_N^{qg}$ is actually a leading effect.
In addition, on adding the quark anomalous dimension
the leading order expression or $F_2(x,Q^2)$
is given by eq.(23), whilst the next-to-leading order result
is more complicated and involves, in particular, terms proportional
to $(\gamma_N^{qg})^2$. Such terms are small in the $\overline{{\mathrm{MS}}}$
scheme but may be large in DIS. The corresponding large terms never appear in a
$\overline{{\mathrm{MS}}}$ scheme calculation since the coefficient function,
$C_{g,N}^{(1)}$, is never iterated (i.e. raised to a power). Consistency
between the schemes therefore suggests a cancellation of large terms in the DIS
scheme. In order to perform a complete next to leading order
calculation of $F_2(x,Q^2)$ there are many other terms which need to be
computed, i.e. the $\sim \alpha_s {\cal{O}}(\alpha_s/N)$ corrections to the
gluon anomalous dimensions $\gamma_N^{gg}$ (and $\gamma_N^{gq}$) and
the higher order corrections to the quark anomalous dimensions, i.e.
$\sim \alpha_s^2 {\cal{O}}(\alpha_s/N)$. Cancellations should then
occur between the terms proportional to the square of the leading order
quark anomalous dimension and these other terms. By not performing the
full next-to-leading calculation, Ellis {\it et al} are unable to
observe these expected cancellations and as such their conclusions, based upon
different assumptions regarding the imposition of momentum conservation, may be
drawn prematurely.

\section*{Conclusions}
We have presented an analytic approach to the evaluation of small $x$ cross
sections and studied the behaviour of the gluon structure function, defined in
a way which is consistent with the previous studies based upon the direct
solution to the BFKL equation. In particular the solution in $x$-space is
obtained exactly, and explicitly reveals the extreme limitations of the double
leading log result and the power behaviour expected from the BFKL approach.
In addition, we examined the deep inelastic
structure functions $F_2(x,Q^2)$ and $F_L(x,Q^2)$ and demonstrated that the
high energy corrections (to the double log calculation) are significant in the
HERA region. Consistency with the data on $F_2(x,Q^2)$ is found. The structure
function $F_L(x,Q^2)$ and the $Q^2$ dependence of $F_2(x,Q^2)$ should be able
to provide sensitive tests of the small $x$ dynamics; in particular deviations
from the traditional approach (expansion in $\alpha_s$) may well be observable.

We have not discussed the process of heavy quark production (in deep inelastic
scattering or in photoproduction), although this process also ought to shed
important light on the essential dynamics \cite{hq}.
Also, the recent measurement by the
ZEUS collaboration of the dijet cross section in photoproduction \cite{zeus}
could be confronted with theory using the techniques presented here
\cite{dijets}. Finally, we wish to make available
the expansion (in $\alpha_s/N$) of the
coefficient function, $R_N$, and the $\overline{ {\mathrm{MS}} }$ scheme
expansions of the quark anomalous dimension $\gamma_N^{qg}$ and the gluon
coefficient function $C_{g,N}^{(1)}$. These are displayed in the following
table.

\section*{Acknowledgments}
We wish to thank Keith Ellis, Francesco Hautmann and Graham Ross
for helpful discussion.

\newpage
\noindent{\large\bf Table}
\begin{center}
Values of the first 18 coefficients in series expansions:
\end{center}

$$
\gamma^{qg (\overline{ {\mathrm{MS}} })}_N = \frac{\alpha_s}{2\pi}
\frac{1}{3} \sum_{n=0}^{\infty} a^{qg}_n \left(
\frac{\bar{\alpha}_s}{N}\right)^n,
$$

$$
C^{(1)(\overline{ {\mathrm{MS}} })}_{g,N} = \frac{\alpha_s}{2\pi}
n_f \frac{1}{3}\sum_{n=0}^{\infty} \tilde p_n \left(
\frac{\bar{\alpha}_s}{N}\right)^n,
$$

$$
R_N^{(\overline{ {\mathrm{MS}} })} = \sum_{n=0}^{\infty} \tilde c_n \left(
\frac{\bar{\alpha}_s}{N}\right)^n,
$$
\begin{center}
i.e. $\tilde p_n = \frac{n_f}{18}p_n$ (see eq.(24)) and
$ \tilde c_n {\bar{\alpha}_s}^n = c_n$ (see eq.(15)).
\end{center}
\vspace*{1cm}

\begin{center}
\begin{tabular}{|r|c|c|c|} \hline
 n & $a^{qg}_n$ & $\tilde p_n$ & $\tilde c_n$ \\ \hline
0 & 1.00 & 1.00 & 1.00 \\
1 & 1.67 & 1.49 & 0.00 \\
2 & 1.56 & 9.71 & 0.00 \\
3 & 3.42 & 1.64$\times 10^1$ & 3.21 \\
4 & 5.51 & 3.91$\times 10^1$ & -0.811 \\
5 & 7.88 & 1.29$\times 10^2$ & 4.56 \\
6 & 2.57$\times 10^1$ & 2.41$\times 10^2$ & 3.27$\times 10^1$ \\
7 & 4.42$\times 10^1$ & 6.53$\times 10^2$ & -2.95 \\
8 & 8.77$\times 10^1$ & 1.93$\times 10^3$ & 1.08$\times 10^2$ \\
9 & 2.83$\times 10^2$ & 4.01$\times 10^3$ & 4.00$\times 10^2$ \\
10 & 5.11$\times 10^2$ & 1.14$\times 10^4$ & 1.33$\times 10^2$ \\
11 & 1.24$\times 10^3$ & 3.17$\times 10^4$ & 2.10$\times 10^3$ \\
12 & 3.68$\times 10^3$ & 7.18$\times 10^4$ & 5.51$\times 10^3$ \\
13 & 7.17$\times 10^3$ & 2.07$\times 10^5$ & 5.30$\times 10^3$ \\
14 & 1.91$\times 10^4$ & 5.52$\times 10^5$ & 3.85$\times 10^4$ \\
15 & 5.29$\times 10^4$ & 1.33$\times 10^6$ & 8.49$\times 10^4$ \\
16 & 1.12$\times 10^5$ & 3.82$\times 10^6$ & 1.40$\times 10^5$ \\
17 & 3.11$\times 10^5$ & 1.00$\times 10^7$ & 6.95$\times 10^5$ \\
\hline
\end{tabular}
\end{center}

\newpage

\noindent{\large\bf Figure Captions}
\begin{itemize}
\item[{[1]}] Contour plot exhibiting the contribution to the full gluon
structure function made by the double leading log term.
\item[{[2]}] Gluon structure function $G(x, Q^2)$ as a function of $x$ plotted
for a range of $Q^2$ values. Contribution made by double leading log
approximation shown by dot-dashed line.
\item[{[3]}] Prediction of longitudinal structure function $F_L(x, Q^2)$
as a function of $x$ plotted for a variety of $Q^2$ values. Contribution made
by double leading log term shown by dot-dashed line. Also shown by
dashed line is the prediction made using the best fit for $F_2(x, Q^2)$
while keeping only the double log term.
\item[{[4]}] Comparison of theoretical predictions with the small $x$
(i.e. $x <0.005$) data from the (a) ZEUS collaboration (renormalised up
2\%) and the (b) H1
collaboration (renormalised down 4\%).
The dotted line corresponds
to the best fit for this expression minus the formally subleading terms
coming from the derivative of the coefficient function and the dashed line to
the best fit for the double leading log approximation.
\end{itemize}

\end{document}